
\input phyzzx
\hsize=6.0in
\vsize=8.9in
\hoffset=0.0in
\voffset=0.0in
\FRONTPAGE
\line{\hfill BROWN-HET-839}
\line{\hfill October 1991}
\vskip1.5truein
\titlestyle{{STRING FIELD ACTIONS FROM $W_{\infty}$}\foot{
Supported in part by the Department of Energy under
contract DE-AC02-76ER03130-Task A}}
\bigskip
\author{Jean AVAN and Antal JEVICKI}
\centerline{{\it Department of Physics}}
\centerline{{\it Brown University, Providence, RI 02912, USA}}
\bigskip
\abstract
\noindent Starting from $W_{\infty}$ as a fundamental symmetry
and using the coadjoint orbit method, we derive an action
for one dimensional strings.  It is shown that on the
simplest nontrivial orbit this gives the single scalar
collective field theory.  On higher orbits one finds
generalized KdV type field theories with increasing number of
components.  Here the tachyon is coupled to higher tensor fields.
\endpage

\noindent {\bf 1. Introduction}
\medskip

Our recent investigations of the collective field formulation of d=1
string theory [1,2] emphasized a number of
characteristic symmetry
properties.  These were described by a  $w_{\infty}$--algebra
of observables which was exhibited first at the classical and then
at the quantum level.  The
theory is described by a scalar field representing the density of
fermions, the chiral components of which obey
decoupled equations arising from the Hamiltonian
$${\cal H} = \int \bigl( {\alpha_+^3\over 3} -
{\alpha_-^3\over 3} \bigr) dx +
\int v (x) \, (\alpha_+ - \alpha_-) dx\eqno\eq$$
with the Poisson structure: $\bigl( u(1)\times u(1)\,{\rm current \,
algebra}\bigr)$
$$\eqalign{ \bigl\{ \alpha_+ , \alpha_+ \bigr\}
& = - \bigl\{ \alpha_- , \alpha_- \bigr\} =
\delta ' (x-y)\cr
\bigl\{ \alpha_+ , \alpha_- \bigr\} & = 0 }\eqno\eq$$
It was shown that the theory was classically Liouville--integrable
for any potential $v(x)$; moreover, due to the Poisson
structure (2), a $(w_{\infty})^2$ algebra structure arose,
generated by
$$h^{\pm}_{\,\ m}{}^{n} \equiv
\int x^{m-1}\, {\alpha_{\pm}^{m-n}\over m-n}\, dx \, , \,
(m\geq n)\eqno\eq$$
and reading
$$\eqalign{\bigl\{ h^{\pm}_{\,\, m}{}^n \, , h^{\pm}_{\,\, p}{}^q \,
\bigr\} & = \bigl( (m-1) q - (n-1) p\bigr) \,\,\,
h_{m+p-2}^{n+q} \cr
\bigl\{ h^+ , h^- \bigr\} & = 0 }\eqno\eq$$
For each chirality we therefore have a $w_{\infty}$ algebra
isomorphic
to the Poisson bracket algebra of polynomials $x^m \, p^{m-n}$.

The $(w_{\infty} )^2$--algebra survives quantization. In the
case when $v(x) = x^2$, it allows the construction [2] of an
exact set of discrete eigen-operators for the Hamiltonian (1).
These were shown to
also realize a  $(w_{\infty})^2$-algebra.  In short the
$(w_{\infty})^2$ algebra was established to be the
spectrum--generating algebra [2,4] for
discrete states of the d=1 string
theory.  A similar conclusion was reached directly from the study
of vertex operators in [5,6].

It appears therefore that the $w_{\infty}$-algebra and
its possible``quantum" deformations such as the algebra of
differential operators
with polynomial coefficients $\bigl\{ x^n \, \partial_x^m \bigr\}$
(of which $w_{\infty}$ is the 1-contraction algebra), plays a basic
symmetry role in any field theory of strings [1-10].  The collective
field theory described above or equivalently the free fermion theory
is just a first example of a more general
structure.

We wish to present here a construction of an extended class of field
theories, using the coadjoint orbit construction [11] applied to the
algebra $w_{\infty}$, using its Lie-Baxter [15] structure.  We will
show that the theory corresponding to the {\it lowest} nontrivial
orbit is precisely the collective tachyon field theory.
This will clarify the Poisson structure (2), the origin of the
Hamiltonian
(1), and the associated representation of the
$w_{\infty}$-algebra (3) which will be understood
as a subalgebra of the algebra of {\it observables} on a symplectic
manifold canonically generated by the coadjoint action of the
algebra of {\it fields} $w_{\infty}$ on its dual $w_{\infty}^*$.

The construction of more general orbits then leads to field theories
with an increasing number of fields.  These fields include the
tachyon, which interacts with the other fields
in a $w_{\infty}$--covariant
way.  In general one has candidate field theories for
any number of these supplementary degrees of freedom.
\medskip

\noindent{\bf 2. The Coadjoint Orbit Construction}
\medskip

It is a well-known feature of dynamical systems having the dual
${\cal G}^*$
of the Lie algebra ${\cal G}$ of a group {G} as phase space, that the
canonical way of obtaining a symplectic manifold in this phase space
(i.e. a bona fide phase space with non-degenerate Poisson brackets)
is by
constructing orbits of the coadjoint action of $G$ on ${\cal
G}^*$ [11].
Representing the action of ${\cal G}^*$ on ${\cal G}$ by the duality
bracket $<>$, the coadjoint action is then given  by:
$$\forall g \in  G , \,\, g^* \in {\cal G}^* ; h\in {\cal G}: \, <
Ad^* h.g^* , g > = <g^* , Ad(h)\cdot g>\eqno\eq$$
where the adjoint action is defined by the Lie bracket:
$$Ad(h)\cdot g = [g, h]\, .\eqno\eq$$
The integrated coadjoint action of the group ${G}$ on an initial
point $b\in{\cal G}^*$ then spans a coadjoint orbit in ${\cal G}^*$.

\noindent We recall the following results:
\item{1.} On each coadjoint orbit there exists a natural
symplectic 2-form, obtained by reduction of the canonical
Kirillov-Poisson bracket
(dual of the Lie bracket) [11].
\item{2.} This symplectic 2-form allows the construction of action
functionals on the coadjoint orbit, exhibiting a global
coadjoint-invariance plus a
local, gauge-type coadjoint invariance group obtained from the
$Ad^*$-invariance group of the original point $b$ in the orbit [12].
They read [13,14]:
$$S = \int d\lambda d\tau \, < Ad^* \bigl( g(\lambda,\tau)\bigr)
\cdot b, \,\, [u_{\lambda}, u_{\tau}]> - \int_{\partial M_1} \,
Hd\tau\eqno\eq$$
where:
\item{.} $b$ is an arbitrarily chosen origin on the orbit ${\cal O}$
\item{.} $g(\lambda , \tau)$ is a 2-parameter element of $G$,
spanning a 2d surface $M_1$ in the orbit ${\cal O}$.
\item{.} $u_{\lambda} , u_\tau$ are elements of ${\cal G}$
respectively describing the coadjoint infinitesimal action of
$g^{-1} \partial_{\lambda} g$ and $g^{-1} \partial_{\tau} g$.
\item{.} $<>$ is the duality bracket, $[\,]$ the Lie bracket.
\item{.} $\tau$ will play the role of time for the Lagrangian (7).
\item{.} $\lambda$ is a supplementary variable, integrated over
whenever possible, allowing to consider topologically non-trivial
phase spaces varying from 0 to 1.
\item{.} $H$ is an arbitrary Hamiltonian function on ${\cal O}$.
\item{.} $\partial M_1$ is the boundary at fixed $\lambda =1$ of
the 2-dimensional surface ${M_1}$ in ${\cal O}$ spanned by
$Ad^* g (\lambda ,
\tau )\cdot \sim Ad^* g (1,\tau)\cdot b$

\item{3.}  The Lagrange equations of motion from the action $S$
in (7) are equivalent to the Hamiltonian equations of motion
generated by $H$ on the boundary $\partial M_1$ [12].

The coadjoint orbit method is crucial in the Lie-Baxter
construction
of integrable systems, which we are now going to describe.
A Lie-Baxter algebra is a Lie algebra endowed with two Lie algebra
structures, namely:
$$\eqalign{ ad(x)\cdot y & = [x,y]\,\, ({\rm usual\,\,
Lie\,\,bracket})\cr
ad_R (x) \cdot y & = [Rx,y] + [x,Ry]\,\, (R-{\rm
commutator}); R \, \in \, {\rm End} ({\cal G}) }\eqno\eq$$
In order for (b) to satisfy the Jacobi identity, $R$ must obey the
classical
Yang-Baxter algebra [15].  In this case, it was shown that

\item{1.} $Ad^*$-invariant functions of ${\cal G}^*$ are in involution
under the two Poisson-Kirillov brackets generated by the two Lie
algebra structures (8) [16].
\item{2.}  The equation of motion on ${\cal G}^*$ (or in fact on a
$Ad_R^*$-coadjoint orbit of ${\cal G}^*$), given by a $Ad^*$-invariant
Hamiltonian (typically a Gelfand-Dikii [17] hamiltonian
$\Tr (g^* )^{n/m}
\vert_{g^*\in S^*}$), has a Lax representation, and is
Liouville--integrable due to 1).

Finally, there exists a well-known method, known as Adler-Kostant-
Symes scheme, to obtain such Lie-Baxter algebras [16].  Whenever the
Lie algebra ${\cal G}$ can be decomposed as a vector space into a
direct sum of two sub-algebras $N$ and $K$, the endomorphism $P_N -
P_K$ (projections on $N, K$ ) is a solution of the Yang-Baxter
equation.  In this case, one can compute the second coadjoint action of
${\cal G}^*_R$ on ${\cal G}_R$:
$$ad_R^* \bigl( n_2 + k_2 \bigr) \cdot \bigl( n_1^* \bigr) =
ad^* (n_2 )\cdot n_1^* + ad^* (k_2 ) \cdot k_1^*\eqno\eq$$
We now apply this formalism to the case when ${\cal G}$ is a
$w_{\infty}$--algebra.
\medskip

\noindent{\bf 3. Coadjoint Orbits in $w_{\infty}^*$-
Algebra}
\smallskip

The $w_{\infty}$--algebra $W$ is generated by the elements $w_m^n$
with the commutator:
$$\bigl[ w_m^n , w_p^q \bigr] = \bigl( (m-1) q - (p-1)n\bigr)
w_{m+p-2}^{n+q}\eqno\eq$$
To any element of $W$ we associate the 2-variable function:
$$h \in w = \sum h_m^n \, w_m^n \rightarrow h(x,y) = \sum h_m^n \,
x^{n+m-1} \, p^{m-1}\eqno\eq$$

The commutator of two elements $h_1 , h_2 $ of $W$ is represented
under the correspondance (11) by the Poisson bracket of $h_1 (x,y) $
with $ \, h_2 (x,y )$, generated by the one--dimensional symplectic
structure $\{ x,y\} = 1$.

The dual algebra $W^*$ is generated by dual elements $w_m^{*n}$ ; the
functional representation reads:
$$\bar{h} \in W^* = \sum \bar{h}_m^n \, w^*_m{}^n \rightarrow \bar{h}
(x,y) = \sum \bar{h}_m^n \, x^{-n-m} \, p^{-m}\eqno\eq$$
so that the duality bracket $< \bar{h} \vert h >$ becomes the contour
integral
$$\oint dx \oint dy \quad \bar{h} (x,y) h (x,y)\, . \eqno\eq$$
from where the coadjoint action immediately becomes, in the functional
representation
$$Ad^* \bigl( h (x,y) \bigr)\cdot \bar{k} (x,y) = \{
\overline{h,\bar{k}} \}\eqno\eq$$
This allows to identify $Ad$ and $Ad^*$ under the identification
$w_m^{*n} = w_{-m+s}^{-n}$.
Now we have an obvious Adler-Kostant-Symes decomposition of $W$ as:
$$\eqalign{ W = W^+ \bigoplus W^- \quad\quad W^+ & = \{ w_m^n , m\geq
1\} \cr
W^- & = \{ w_m^n , m\leq 0\} }\eqno\eq$$

The Lie-Baxter algebra structure reads:
$$\bigl[ w_1^+ + w_1^- , w_2^+ + w_2^-\bigr] = \bigl[ w_1^+ , w_i^+
\bigr] + \bigl[ w_1^- , w_2^- \bigr] \eqno\eq$$
The identification of $W$ and $W^*$ becomes an identification between
$(W^+ )^*$ and $W^-$, and $(W^- )^*$ and $W^+$.  The coadjoint action
then reads:
$$Ad_R^* \, \bigl( w_1^+ + w_1^- \bigr) \cdot \bigl( w^+  +
w^- \bigr) = \bigl[ w_1^+ , w^- \bigr]_- + \bigl[ w_1^- ,
w^+\bigr]_+\eqno\eq$$
We shall now restrict ourselves to coadjoint orbits in $W^+ \simeq
(W^- )^*$.  The coadjoint action (14) reduces to the projected
(standard) coadjoint action of $W^- \sim \bigl[ W^- , W^+ \bigr]_+$.

Let us first describe in detail the simplest non--trivial
example of a coadjoint orbit. It turns out to
correspond to the string collective field theory [1,2].  Consider the
subset of $(W^- )^*$ parametrized by the functions:
$$h(x,y) = y^2 + u (x)\eqno\eq$$
$u(x)$ being any function of $x$, formally $u = {\sum\atop n\in
{\bf Z}} \, u^{(n)} x^n$.

We now apply the construction scheme described in Section 2:

(a)  The set $h(x,y)$ parametrizes a coadjoint orbit of $W^+$.
Indeed, the coadjoint action of $W^-$ under (14) reads:
$$ Ad^* \, (W^- )\cdot h  = \bigl\{ y^2 + u(x) , y^{-1} v_1 (x)
+ y^{-2} v_2 (x) + \cdots \bigr\}_+ \, = -2\,
{\partial v_1\over\partial x}\eqno\eq$$
We see that $u (x)$ transforms into $u(x) - 2v_{1,x}$,
hence any function $u(x)$ can be reached from a given $u_0 (x)$
provided that $\int dx \, u_0 (x) = \int u(x) $.  We recover also the
argument that any integral of the form $\int dx \,
{\partial v\over \partial x} \equiv 0$ for any
meromorphic function $v(x)$; this argument was used [1,2] in our
derivation of the $w_{\infty}$ - algebra.

(b)  The representatives of $g^{-1} \partial_{\tau} g$ and $g^{-1}
\partial_{\lambda} g \, \in W^-$, defined in (7), can be explicitely
evaluated.  Indeed, the coadjoint action simply translates into a
``deformation" of $u(x)$ as $u(x,\lambda , \tau)$.  In particular
it follows from (19) that
$${\partial u\over \partial \tau ,\lambda} = - 2\,
{\partial v_1^{\tau , \lambda}\over \partial x}\eqno\eq$$
where $g^{-1} \partial_{\tau ,\lambda} g \sim y^{-1} v_1^{
(\tau,\lambda )} \, (x) + \cdots$  Hence, inside (7):
$$\eqalign{ u_{\tau} & \equiv {1\over 2} y^{-1} \,
\partial_x^{-1} \, {\partial u\over \partial\tau} + \cdots\cr
u_{\lambda} & \equiv {1\over 2} y^{-1} \, \partial_x^{-1} \,
{\partial u\over\partial\lambda } + \cdots\cr
Ad_R^* \, g\cdot b & \equiv y^2 + u (x, \lambda \tau ) }\eqno\eq$$

(c)  The action $S$ (7) therefore becomes:

$$S  = \int dx \, d\lambda dt\,  {1\over 4}\, \bigl(
\partial_x^{-1} \, u_{\tau}\, u_{\lambda} - \partial_x^{-1} \,
u_{\lambda} \, u_{\tau}\bigr) \, \, - H(u)\, \   $$
Introducing the 1+1 dimensional field
$u(x,\tau ) \equiv u (x,\lambda = 1 , \tau )$ we have after
integration
$$ S = - {1\over 4} \int dx \, dt \, u \, \partial_x^{-1} \,
{du\over dt} - H(u)  \, .\eqno\eq$$

{}From (22), the canonical Poisson structure of any field theory on
this orbit is easily read:
$$\Pi_u \equiv {\partial{\cal L}\over \partial\dot{u}} =
{1\over 4}\,\partial_x^{-1} u \Rightarrow \{ u(x) , u(y)\} =  4\delta '
(x-y)\eqno\eq$$

We have recovered the Poisson structure (2) of the collective field
theory.  Now it follows from the Lie-Baxter algebra construction that
there exists a set of commuting Hamiltonians on the $Ad_R^*$-invariant
orbits, i.e. the $Ad^*$-invariant functions.  Such functions are
easily obtained, once one introduces the $Ad^*$-invariant trace:
$$\Tr (w\in W) = \int dx dy w(x,y)\eqno\eq$$
Its crucial $Ad^*$-invariance properly follows from:
$$\Tr \bigl( [w_1 , w_2 ]\bigr) = \int dxdy \, \{ w_1 , w_2\}
= 0 \,\, {\rm by\,\, part\,\, integration}\,.$$
It can therefore be used to define {\it $Ad^*$}-invariant
functions of $W^+$, namely the equivalent of Gelfand-Dikii
hamiltonians [17].  $\Tr \, L^{p/n} $ for $L \sim y^n + \cdots$
in $W^{\infty}$.  In
particular, on the orbit in $W^+ \sim y^2 + u (x)$, the
$Ad^*$-invariant functions reduce to:
$$I_k = \Tr \, \Bigl( \bigl[ y^2 + u(x) \bigr]^{2k+1/2} \bigr) \,
 \sim \int u^{k+1} (x) dx\eqno\eq$$
In this way we recover naturally the hierarchy of commuting
Hamiltonians for the zero-potential case.

The first nontrivial Hamiltonian with $k=2$ gives the cubic
interaction and the Lagrangian
$${\cal L} = {1\over 4} \bigl(u_+ \, \partial_x^{-1} \, \dot{u}_+ -
u_- \, \partial_x^{-1} \, \dot{u}_- \bigr) - {1\over 12\sqrt{2}} \,
\bigl( u_+^3 - u_-^3 \bigr)\eqno\eq$$
where we have added an identical $w_{\infty}$--coadjoint action for
the other chirality component.  Defining $u^{\pm} = \sqrt{2} \bigl(
\Pi_{,x} \pm \phi\bigr) $ one ends up with the cubic collective field
theory.
$${\cal L}_{\rm coll} \, = \Pi \, \dot{\phi} - \bigl( {1\over 2} \,
\Pi_{,x} \, \phi \, \Pi_{,x} + {1\over 6} \, \phi^3 \bigr)\eqno\eq$$
Let us now comment on the Poisson structure (23).  It is easy to check
the consistency of the procedure by comparing this structure with the
restriction to the orbit of the Kirillov-Poisson structure
(for $W$ or
$W_R$, since there is no difference when one considers solely the
subalgebra $W^+$).  The $KP$ bracket on $W_-^*$ reads [3],
$$\bigl\{ h_m^{*n} , \, h_p^{*q} \, \bigr\} = \bigl((p-1) n - (m-
1)q\bigr) \, h_{m+p-2}^{*n+q}\eqno\eq$$
Reduced to the orbit $y^2 + u(x)$, which corresponds to the constraint
$\{ h_2^{-2} = 1, h_m^p = 0 \, \forall m \not= 0,2\}$, it gives:
$$\bigl\{ h_0^n , h_0^m \bigr\} = (m-n) \, h_{-2}^{m+n} = (m-n) \,
\delta_{m+n ,2}\eqno\eq$$
thereby identifying the canonically conjugate variables on the orbit
as $\{h_0^n , h_0^{2-n}$ for $n>1\}$ , plus the invariant $h_0^1 \sim
\int u\, dx$.  The coordinates on the orbit $h_0^{*n}$ are the
coeffients of the formal expansion of $u$ as $u = \sum h_0^{*n} \,
x^{-n}$, or $h_0^{*n} = \oint x^{n-1} u(x) dx$.  This is a
very simple realization of the Darboux
theorem for this symplectic manifold.  From (29) and using:
$$\sum_{{n\geq 0\atop \leq 0}} \, n\, x^{-n-1} \, y^{n-1} = \Delta
(x,y) = \delta ' (x-y) \quad ({\rm under}\, \oint \, {\rm
integration})\eqno\eq$$
the Poisson structure (23) follows.

The orbit is now explicitely described as a direct sum of
${\bf N}$--labeled phase spaces, therefore
the algebra of functions on the orbit or, loosely speaking, the
algebra of observables in the corresponding quantum theory,
is a ``product" of $w_{\infty}$--algebras, precisely,
$${\cal W} = \bigoplus_{n=1}^{\infty} \, \Bigl\{ \bigotimes_{i=1}^n
\,  w_{(i)} \Bigr\}\eqno\eq$$
where
$$ w_{(i)} = \bigl\{ x_{(i)}^m \, p_{(i)}^n \,m, n \in {\bf Z}
\bigr\}\eqno\eq$$
denotes the  $w_{\infty}$ algebra associated to the i$^{th}$
phase space.  We recognize here the structure of the matrix model
where the eigenvalues $\lambda_i$ and their conjugates $p_i$
generate a product of $w_{\infty}$ algebras.
Our field--theoretical $(w_{\infty})^2$-algebra
$\int x^m \, \alpha_{\pm}^n$ is now understood as being
a subalgebra of this large ${\cal W}$-algebra.
${\cal W}$ would in
fact be described by {\it all} generators of the form:  $\bigl\{ \int
x^m \, \alpha^{n_1} \, \alpha^{\prime n_2} \, \cdots \alpha^{(p)n_p}
, p = 1 \cdots \infty \bigr\}$.  The ``fundamental" algebra or
algebra of fields is in fact the
original $w_{\infty}$ algebra.  Note that this construction indicates
that the ``$\int dx$" symbol in the construction of the algebra
$\bigl( \int x^m \, \alpha^{n-m}\bigr)$ must be understood as
$\oint$, and it was therefore allowed to suppress all integrated terms.
There is one exception, however, when one computes
$$\bigl\{ \, \int x^n \alpha , \int x^{-n} \alpha\bigr\} = \int x^n\,
y^{-n}\, \delta ' \, dxdy = \int x^{-1} dx \equiv 1\eqno\eq$$
(and not $0$, but $x^{-1} = \partial_x^{-1} \, \ell nx$ and $\ell nx$
is not a meromorphic function)\nextline
Any other commutator, however, was evaluated in [1,2] using formal
partial integrations, whenever the final order in $\alpha$ was non--
zero (which is always consistent with the definition of $\int dx$ as
$\oint$, except in the above case, when one
considers generators with a \underbar{positive} power of $\alpha$),
and is therefore correctly given by the standard (4)-formula.

We have now exhibited the structure of the collective string theory
in $d=1$ as a coadjoint construction on the algebra $w_{\infty}$.
We have found that it is given by the lowest orbit, defined as
$y^2 + u(x)$. It is clearly interesting to study the more general
orbits, which lead to increasing number of fields and
the extension of the simple tachyon collective field theory.
\medskip

\noindent {\bf 4.  High-Dimensional Orbits and Extended
Lagrangians}
\medskip
The higher--dimensional orbits follow from the same type of
construction which we have extensively described in the lowest--
dimensional case.  We shall now describe the main
characteristic features and give some typical examples for
Lagrangian field theory associated with higher $w_{\infty}$ hierarchy.

First of all, a set in $W_-^*$ represented by functions of the form
$F_n (y,z) = y^n + u^{(n-2)} \, y^{n-2} + \cdots u^{(0)}$ is invariant
under $ad^* (W_- )$.  One ignores the possible term $u^{(n-1)} \,
y^{n-1}$ since it is left invariant by the coadjoint action and is
therefore irrelevant in the description of a coadjoint orbit.  The
above manifold, with arbitrary functions $u^{(n-2)} \cdots u^{(0)}$
(i.e. arbitrary coordinates $\bigl( h_{-n-2}^{\cdots} \cdots
h_0^{\cdots} \bigr)$ on the basis $w_{-n}^{*m}$ of $W_-^*$ ),
is a coadjoint orbit
(up to a finite number of degrees of freedom corresponding to
normalization constants $\sim \int u^{(p)}$ ).  Indeed, the coadjoint
action reads:
$$\eqalign{ ad^* W_- \cdot \, F_n & = \bigl\{ y^n + u^{(n-2)} \,
y^{n-2} + \cdots
u^{(0)} , y^{-1} b^{(1)} + \cdots y^{-n+1} \, b^{(n-1)} + \cdots
\bigr\} \vert_+\cr
& = ny^{n-2} \, {\partial b^{(1)}\over \partial x} + n y^{n-3} \,
{\partial b^{(2)}\over\partial x} + \cdots \, \bigl( n\, {\partial
b^{(n-1)}\over\partial x} + \cdots 2 \, {\partial u^{(2)}\over
\partial x} \, b^{(1)}\bigr) }\eqno\eq$$
hence the form of the function $F_n$ is $ad^*$--invariant, and all
functions $u^{(n)}$ can be reached by $ad^*$--action from arbitrary
initial functions $u_0^{(n)}$, provided that normalization
invariances such as $\int u_0^{(n-2)} = \int u^{(n-2)} \cdots$ be
respected.The Poisson bracket structure on this orbit (which is a
symplectic structure up to a finite number of central degrees
of freedom) is readily obtained from the reduction of the
Kirillov-Poisson structure of $W_-^*$, given in (28), to  the orbit
$\bigl\{ h_{-n}^n = 1 , \,\, h_{-p}^m = 0$ for $p \not= 0
\cdots -n-2\bigr\}$.  Using (28) together with:
$$\sum_{n\in{\bf Z}} \, x^{-n-1} \, y^n \sim \delta (x-y) \quad ({\rm
under} \,\, \oint )\eqno\eq$$
leads to the following Poisson structure:
$$\eqalign{ & \bigl\{ u^{(p)} (x), u^{(q)} (y) \bigr\} =
{1\over 2} (p-q) \, \delta (x-y) u_{,x}^{(p+q+2)} (x) + \cr
& + {p+q+2\over 2} \,
\delta_x^{\prime} (x-y) \cdot \Bigl(  u^{(p + q + 2)} \, (x) +
u^{(p+q+2)} \, (y)\Bigr) }\eqno\eq$$

Although it does not seem obvious to write immediately a Lagrangian
leading to such a Poisson structure, we know in principle how to do
so:  one has to ``solve" formally (34) in order to express $b^{(\tau ,
\lambda )}$ in terms of $\partial u^{(p)}/\partial\tau$ and
$\partial u^{(p)}/\partial\lambda$.  Obviously, this can be achieved
in a technically simple, although becoming tedious when $n$ is large,
recursion scheme.  Once explicitely obtained, one plugs back
$u_{\tau} $ and $u_{\lambda}$ into (7), and obtains the explicit
Lagrangian.  We expect however that drastic simplifications occur
when the fields $u^{(n)}$ are suitably redefined.  Although we have no
general proof, we have shown up to under $n=5$ (4 distinct functions
$a$) that (36) can be recast as: (for instance, $n=5$ )
$$\bigl\{ \bar{u}^{(0)} , \bar{u}^{(3)} \bigr\} = 5 \, \delta ' (x-y)
\quad \bigl\{ \bar{u}^{(1)} , \bar{u}^{(2)} \bigr\} = 5 \, \delta '
(x-y)\eqno\eq$$
all other brackets vanish.\nextline
This follows from redefining in (36):
$$\bar{u}^{(0)} = u^{(0)} - {1\over 5} \, u^{(2)} \, u^{(3)} \quad
\bar{u}^{(1)} = u^{(1)} - {3\over 20} \, u^{(3)^2}\eqno\eq$$
The Lagrangian obtained from (7) then becomes
$${\cal L}  = \int  dx \quad \bigl( \dot{u}^{(0)} \, \partial_x^{-1}
\, u^{(3)} \, + \, \dot{u}^{(1)} \, \partial_x^{-1}\, u^{(2)}\bigr) -
H \eqno\eq$$
If our conjecture is correct, for any $n$ one can redefine the
$u^{(p)}$ functions in (36) so as to reduce it to
$$\bigl\{ u^{(p)} , \, u^{(n-p-2)} \bigr\} \cong \delta ' (x-y) \quad
\forall \, p = 0 \cdots (n-2)\eqno\eq$$
In particular, for $n$  even, the field $u^{({n-2\over 2})}$ is
``self-conjugate"  under (40), just as the tachyon field $\alpha$ was
for $n=2$.  It seems natural to try an identification of this field as
the tachyon field, coupled to the other fields $u^{(p)}$ and their
``momenta" $u^{(n-p-2)}$.

The couplings are described by a specific choice of Hamiltonian.  We
know in principle an infinite hierarchy of commuting Hamiltonians
which would lead us to integrable field theories.  They read (for the
$n^{th}$ hierarchy), redefining the labeling of $u$ as
$u^{(p)}\rightarrow u^{(n-p)}$
$$H^p = \int \bigl[ y^n + u^{(2)} \, y^{n-2} + \cdots
u^{(n)}\bigr]_+^{p/n} \, dx\,dy\eqno\eq$$
The explicit computation is a technical matter which we shall not
discuss in detail.  The terms that can be obtained are deduced from
the following rules:
(a) the $1/n$--power of $L\equiv y^n + \cdots u^{(0)}$ has the
generic form
$$L^{1/n} = y + \sum_{q=1}^{\infty} \, \sum_{\{p_j\}}
\, {\prod_j a^{(p_j )}\over y^q} \eqno\eq$$
when for a fixed $q$ the set $\{p_j\}$ is such that $\sum p_j = q+1$
and $u^{(0)} = 1 , u^{(1)} = 0$

\noindent (b) it follows that the contributing terms to (41)
of global order $y^{-1}$ are necessarily of the form $\prod_j\,
u^{(p_j)}$ such that $\sum p_j = p+1$  Let us now describe
Hamiltonians and Lagrangians for the next orbits $(n=3 , n=4)$.
The $n=3$ orbit $y^3 + uy + v$ leads to the following
Hamiltonians.
$$\eqalign{ L^{1/3} \rightarrow \int u \, dx \quad\quad\quad & L^{2/3}
\rightarrow \int v \, dx\cr
L^{4/3} \rightarrow \int uv\,dx \quad\quad & L^{5/3} \rightarrow
\int 9v^2 - u^3 dx \cdots }\eqno\eq$$
The associated Lagrangian takes the form:
$${\cal L}  = \int \bigl( \dot{u} \, \partial_x^{-1} v + g_1
\cdot uv + g_2 \bigl( 9v^2 - u^3\bigr) + \cdots \bigr) dx \eqno\eq$$
The $n=4$ orbit $y^4 + uy^2 + vy + w$ leads to the following
Hamiltonians:
$$\eqalign{ L^{1/4} \rightarrow & \int u\, dx \quad\quad L^{1/2}
\rightarrow \int v\,dx \quad\quad L^{3/4} \rightarrow \int \bigl(
8w-u^2 \bigr) dx\cr
L^{5/4} \rightarrow & \int - 3u^3 + 2uw +v  }\eqno\eq$$
and associated Lagrangian:
$$\eqalign{ & {\cal L} =  \bigl( \dot{u} \, \partial_x^{-1} w +
\dot{w} \, \partial_x^{-1} u \bigr) + \dot{v} \,
\partial_x^{-1} v\cr
& + g_1 \, \bigl( 8w - u^2 \bigr) + g_2 \bigl( -3u^3 + 2uw + v^2
\bigr) + \cdots } \eqno\eq$$

In this way, one shall obtain integrable, $w_{\infty}$--invariant
field theories which can be interpreted (for even $n$) as describing
the coupling of the tachyon field $u^{({n\over 2}-1)}$ to other fields
in a very naturally-defined way.  The exact interpretation of these
fields will be left for forthcoming studies;
one must however underline their resemblance with
the fields associated to the hierarchy of discrete states in the
formalism of Klebanov and Polyakov [6].  Our sequence of fields in
the $n\rightarrow \infty$ limit: ${\sum\atop m\geq 0} \, u^{(m)} \,
\bigl( x,\tau\bigr) y^m = U (x,y,\tau )$ becomes a 2+1 field theory.
However one must be
careful about taking the limit
$n\rightarrow\infty$.   This limit
would indeed correspond
to taking the {\it full} $W_+^{\infty}$ algebra as phase space,
instead of its finite $n$ orbit, and the Poisson structure is
degenerate on this large phase space, preventing the construction
of a Lagrangian on the previously described lines although it now
reminds us of gauge--like theories.  This point clearly deserves
further investigations.

Finally we wish to comment on another possible extension of the
previous coadjoint construction.  As indicated above, $W^{\infty}$ is
the 1-contraction reduction of the algebra of pseudo differential
operators.  The above constructions have therefore the simple
interpretation of long wave--length limits of the generalized KdV
hierarchy obtained by application of the AKS scheme to
$Ps$Diff [18].  In fact, the phase space interpretation of $Ps$Diff
was emphasized recently by Yoneya [10].
In Wigner representation, $Ps$Diff operators can be represented by
functions of two variables $x$ and $\zeta$ , of the form $F =
{\sum\atop n\in{\bf z}} \, u^{(n)} (x) \zeta^n$, where the
coefficients $u_p^{(n)} = \oint\,x^{-p-1}\,u^{(n)}$ are the coordinates
on the basis $(\sim x^n \partial_x^p )$ of a given element in $Ps$Diff.
The algebra is described by the product rule:
$$F_1 \cdot F_2 = \sum_{n\geq 0} \, \bigl(
{\partial\over\partial\zeta}\bigr)^n \, F_1 \cdot \bigl(
{\partial\over\partial x}\bigr)^n \, F_2\eqno\eq$$
$W_{\infty}$ would be obtained by restricting $n$ to $1$.  One then
proceeds in defining the Lie--Baxter structure of $Ps$Diff as
$ \bigl\{ {\sum\atop n\geq 0} \, u^{(n)} (x) \zeta^n\bigr\} \oplus
\bigl\{ {\sum\atop n<0} \, u^{(n)} (x) \zeta^n\bigr\}$, the
duality bracket $<F_1 , F_2 > =
\Tr \, F_1 \cdot F_2$ and the adjoint--invariant trace $\Tr F = \oint
\, d\zeta\, dx\,F$.  Finally the Lagrangian on a given coadjoint orbit of
order $n$ in $\zeta$ reads:
$${\cal L} = \oint \, d\zeta\, \oint\, dx \,\, \bigl( \zeta^n + u^{(n-
2)}\zeta^{n-2} + \cdots \bigr) \cdot \bigl[ a_{\tau} , a_{\lambda}
\bigr] \, d\tau \lambda\eqno\eq$$
when $a_{\tau ,\lambda} (x,y)$ is such that:
\item{.} $a_{\tau ,\lambda} (x,y)$ in ($Ps$
Diff) $= {\sum\atop n\geq 1} \, \zeta^{-n} \,
a_{\tau,\lambda}^{(n)}$ is defined by the equation
$$\bigl[ \zeta^n + u^{(n-2)}\zeta^{n-2} + \cdots , a_{\tau
,\lambda}\bigr] \vert_+ = {\partial u^{(n-2)}\over\partial\tau ,
\lambda} \, \zeta^{n-z} + \cdots\eqno\eq$$

. [] is the initial Lie bracket, $[]_+$ is the projection on
($Ps$Diff)$_+$ corresponding to taking the second Lie-Baxter
coadjoint action of ($Ps$Diff)$_-$ on $Ps$Diff$_+$ .
All these features are well-known in the theory of
generalized KdV hierarchies.

We want to emphasize now the relation with
matrix models and fermions.  As before, the algebra of
observables is now an infinite number of copies of operators of
the type $X^n\,\partial^m $:
$$ X_i^n \, \partial_i^m \quad \quad i = 1,2 \cdots \eqno\eq$$
\bigskip

\noindent{\bf 5.  The Free Fermion Formalism}

It is well known [21] that classical integrable equations obtained
from the coadjoint construction on the $Ps$ Diff algebra can be
reformulated in terms of integral equations obeyed by free fermion
correlation functions.  Since this relation may help to understand the
meaning of the supplementary fields $u^{(n)}$ introduced in section 4,
by reformulating them in terms of free fermions (which are known to be
an alternative description of matrix models [19]), we shall describe
briefly its salient points:

One starts from free fermion generators$\bigl\{ \psi_m^+ , \psi_n ,
m,n \in {\bf Z}\bigr\}$, constructing the $Gl (\infty)$ algebra as
bilinears in $\psi$:  $g_{mn} \equiv \psi_m^+ \, \psi_n$, with
commutation relations $[g_{mn} , g_{pq} ] = \delta_{np} \, g_{mq} -
\delta_{mq} \, g_{pn}$.  One introduces a hierarchy of commuting
Hamiltonians $\bigl\{ H^{(n)} = \sum_{i\epsilon{\bf Z}} \, \psi_i^+ \,
\psi_{i+n}\bigr\} , n\geq 0$, and it is easy to show, by a
bosonization argument, that the evolution of the free fermions is
encapsulated in the following form.
$$\eqalign{ \psi(k) = \sum_{n\epsilon{\bf Z}} \, k^{-n} \, \psi_n
\quad ; &  \quad e^{\Sigma x_nH^{(n)}} \cdot \psi \cdot
e^{-\Sigma x_n H^{(n)}} = e^{\zeta (x,k)} \, \psi (k)\cr
& \zeta (x,k) = \sum_{n>0} \, x_n \,
k^n }\eqno\eq$$

Indeed the bosonization formula read:
$$\alpha (k) = \psi^+ (k) \psi (k) \quad ; \quad \psi (k) =
e^{\partial^{-1} \alpha (k)}\eqno\eq$$
and the Hamiltonians $H^{(n)}$ are expressed as:
$$H(x) = \sum \, x_n \, H^{(n)} = \sum \, x_n \, \alpha^{(n)}
\, ; \, \alpha^{(n)} = \int k^{n-1} \alpha (k) dk\eqno\eq$$
from which (51) immediately follows.  The major point consists in
computing particular correlation functions:
$$\eqalign{\tau (x\cdots ) (g) &= < 0 \vert \, e^{-H(x)} \,
g \vert 0 >\,,  \quad g \, \in \, Gl(\infty)\cr
& \vert 0 > {\rm being \,\, the\,\, Fermi\,\, sea\,\,vacuum}
}.\eqno\eq$$
These correlation functions or ``tau functions" obey a particular
bilinear equation or ``Hirota equation" [22]:
$$\oint \, e^{\zeta (x-x_,' k)} \, \tau \bigl(x_n -{1\over nk^n} \cdots
\bigr) \, \tau \bigl( x_n' + {1\over nk^n} \cdots \bigr) dk =
0\eqno\eq$$
from which the differential equations of the KP
hierarchy are obtained by Taylor-expanding (55) and introducing
functions $\sim u = {\partial^2\over \partial x_i^2} \, \ln \tau$.
The KdV hierarchy is obtained in a similar way by a reduction of
$Gl(\infty)$ to the Kac-Moody algebra $A_1^{(1)}$, or equivalently by
suppressing dependence in the even variables $x^{(2n)}$ in (55).  It
follows that the classical fields $u^{(n)} (x,t)$ introduced in section
4 are interpreted as (derivatives of) correlation functions for
particular operators in a free-fermion theory.  Notice also that the
bilinear operators:
$$\Theta^{(n,m)} = \int dk \, \psi^+ (k) \, k^n \partial_k^m \,
\psi (k)\eqno\eq$$
naturally realize the $Ps$ Diff algebra, thereby providing us with an
algebraic relationship between the two formalisms.  In relation to
these remarks, let us finally emphasize that the $Ps$ Diff algebra,
and indeed the whole KP-construction framework, was used in [10,20] to
formulate effective actions for string equations,
implying a possible interpretation of the higher orbit coadjoint
action as an effective action for fermion correlation function.

\noindent {\bf References}

\medskip

\pointbegin
J. Avan and A. Jevicki, {\it Phys. Lett} {\bf B266} (1991), 35.
\point
a)  J. Avan and A. Jevicki, to appear in {\it Phys. Lett.} {\bf B}
 (Brown HET-
821).\nextline
b)  J. Avan, Proceedings of Carg\`ese Summer School on Field Theory,
(1991), edited by G. Mack, to appear (Plenum).
\point
D. Minic, J. Polchinski and Z. Yang, UT preprint UTTG-16-9l.
\point
G. Moore and N. Seiberg, Rutgers and Yale preprints RU-91-29.
\point
E. Witten, IASSNS-HEP-91151.
\point
I. Klebanov and A. Polyakov, Princeton preprint PUPT-1281.
\point
D. J. Gross, I. Klebanov and M. J. Newman, Nucl. Phys. {\bf B350},
(1991) 621.
\point
M. Awada and S. Sin, UFIFT-HEP-91-03 (1991); HFIFT-HEP-90-33
(1990).\nextline
H. Itoyama and T. Matsuo, {\it Phys. Lett. } {\bf B262} (1991), 233.
\point
S. Das, A Dhar, G. Mandal and S. Wadia, ETH-TH-91-30 (1991).
\point
T. Yoneya, UT-Komaba 91-8 (1991).\nextline
R. Dijkgraff, E. Verlinde and H. Verlinde, {\it Nucl. Phys.}, {\bf
B348} (1991), 435.\nextline
M. Fukuma, K. Kawai and R. Nakayama, KEK-TH-272 (1991)
\point
A. A. Kirillov,``Elements of the Theory of Representations"
 (1976).
\point
F. Zaccaria et al., {\it Phys. Rev.} {\bf D27} (1983), 2327.
\point
A. Alekseev, and S. Shatashvili, {\it Nucl. Phys.} {\bf B 323} (1989),
719.\nextline
A. Alekseev, L. Faddeev and S. Shatashvili, {\it J. Geom. Phys.} {\bf
1} (1989), 3.
\point G. Delius, P. van Nieuwenhuizen and V. G. Rodgers, {\it Int.
Jour. Mod. Phys.}, {\bf A5} (1990), 3943.
\point
M. Semenov-Tjan-Shanskii, {\it Funct. Anal. Appl.} {\bf 17} (1983),
17.
\point
M. Adler, {\it Inv. Mat.} {\bf 50} (1979), 219.\nextline
A. G. Reyman, M. Semenov-Tjan-Shanskii, {\it Inv. Mat.} {\bf 54}
(1979), 81; {\bf 63} (1981), 423.\nextline
W. Symes, {\it Inv. Mat.} {\bf 59} (1980), 13.\nextline
B. Kostant, {\it Adv. Mat.} {\bf 34} (1979), 195.
\point
I. M. Gelfand and L. A. Dikii, {\it Func. Anal. Prilozh.} {\bf 10}
(1976), 13.\nextline
V. G. Drinfeld and V. V. Sokolov, {\it Journ. Sov. Math.}, {\bf 20}
(1985), 1975.
\point
G. Segal, G. Wilson,  Publ. IHES {\bf 61} (1985), 5.
\point E. Br\'ezin, C. Itzykson, G. Parisi and J. B. Zuber, {\it Comm.
Math. Phys.} {\bf 59} (1978), 35.
\point
A. Jevicki and T. Yoneya, {\it Mod. Phys. Lett}, {\bf A5} (1990), 1615.
\point
M. Jimbo, T. Miwa, R.I.M.S. Publ {\bf 19} (1983), 943.\nextline
M. Sato, Y. Sato, R.I.M.S. Kokyuroku {\bf 388} (1990), 183; {\bf 414}
(1981), 181.
\point
R. Hirota, in ``Solitons", edited by R. K. Bullough and P. J. Caudrey,
Springer (1980).

\end